\documentclass[aps,prx,twocolumn,superscriptaddress,longbibliography]{revtex4-1}
\usepackage{graphicx}
\usepackage{color}
\usepackage{amsfonts}
\usepackage{ulem}

\definecolor{brown}{RGB}{200,100,0}

\def\kbar{$\bar{\mathrm{K}}$}
\def\mbar{$\bar{\mathrm{M}}$}
\def\kbarp{\kbar$^{\prime}$}
\def\mbarp{\mbar$^{\prime}$}
\def\gbar{$\bar{\mathrm{\Gamma}}$}
\def\kgk{\kbar-\gbar-\kbarp}
\def\mgm{\mbar-\gbar-\mbarp}
\def\gkm{\gbar-\kbar-\mbar}
\def\WS2{WS$_2$}
\def\MoS2{MoS$_2$}{

\begin{document}
\title{Electronic Structure of Exfoliated and Epitaxial Hexagonal Boron Nitride}
\author{ Roland J. Koch}
\affiliation{ Advanced Light Source, E. O. Lawrence Berkeley National Laboratory, Berkeley, California 94720, USA}
\author{ Jyoti Katoch}
\affiliation{ Department of Physics, The Ohio State University, Columbus, Ohio 43210, USA}
\author{ Simon Moser}
\affiliation{ Advanced Light Source, E. O. Lawrence Berkeley National Laboratory, Berkeley, California 94720, USA}
\author{Daniel Schwarz}
\affiliation{ Advanced Light Source, E. O. Lawrence Berkeley National Laboratory, Berkeley, California 94720, USA}
\author{ Roland K. Kawakami }
\affiliation{ Department of Physics, The Ohio State University, Columbus, Ohio 43210, USA}
\author{ Aaron Bostwick}
\affiliation{ Advanced Light Source, E. O. Lawrence Berkeley National Laboratory, Berkeley, California 94720, USA}
\author{ Eli Rotenberg}
\affiliation{ Advanced Light Source, E. O. Lawrence Berkeley National Laboratory, Berkeley, California 94720, USA}
\author{ Chris Jozwiak}
\affiliation{ Advanced Light Source, E. O. Lawrence Berkeley National Laboratory, Berkeley, California 94720, USA} 
\author{ S{\o}ren Ulstrup}
\affiliation{ Advanced Light Source, E. O. Lawrence Berkeley National Laboratory, Berkeley, California 94720, USA}
\affiliation{Department of Physics and Astronomy, Interdisciplinary Nanoscience Center (iNANO), Aarhus University, 8000 Aarhus C, Denmark}
\affiliation{Address correspondence to ulstrup@phys.au.dk}

\begin{abstract}
Hexagonal boron nitride (hBN) is an essential component in van der Waals heterostructures as it provides high quality and weakly interacting interfaces that preserve the electronic properties of adjacent materials. While exfoliated flakes of hBN have been extensively studied using electron transport and optical probes, detailed experimental measurements of the energy- and momentum-dependent electronic excitation spectrum are lacking. Here, we directly determine the full valence band (VB) electronic structure of micron-sized exfoliated flakes of hBN using angle-resolved photoemission spectroscopy with micrometer spatial resolution. We identify the $\pi$- and $\sigma$-band dispersions, the hBN stacking order and determine a total VB bandwidth of 19.4~eV. We compare these results with electronic structure data for epitaxial hBN on graphene on silicon carbide grown \textit{in situ} using a borazine precursor. The epitaxial growth and electronic properties are investigated using photoemission electron microscopy. Our measurements show that the fundamental electronic properties of hBN are highly dependent on the fabrication strategy. 
\end{abstract}

\maketitle

\section{Introduction}

The insulating two-dimensional (2D) material hexagonal boron nitride (hBN) is an important building block in van der Waals heterostructures as it provides dielectric separation while maintaining an atomically sharp interface that is chemcially inert \cite{Geim:2013,Wang2013,Li2016,wang:2017}. Using hBN as a substrate \cite{Dean:2010} instead of SiO$_2$ \cite{Chen:2008} has dramatically improved the transport properties of graphene devices \cite{Meric:2013}, and unveiled important many-body physics \cite{Gorbachev:2012,Young:2012}. Heterostructures of graphene and hBN (G/hBN) have been successfully employed to demonstrate quantum transport phenomena \cite{Mishchenko:2014} with unique quantum fractals and Dirac cone replicas resulting from the G/hBN superlattice \cite{Dean:2013,Yang2013,Wang:2016b}. The structural-, electronic- and optical properties of hBN play an important role for these observations. Similarly to graphene, the properties of hBN are completely determined by a $\pi$-electron system and sp$^2$ covalent B-N bonds organized in a honeycomb structure, but with a wide band gap between the valence band (VB) maximum and conduction band (CB) minimum. In addition to providing an ideal interface with other 2D materials, this simple electronic structure also gives rise to deep ultraviolet emission in bulk hBN \cite{Watanabe:2004} and room-temperature quantum emission in single-layer hBN \cite{Tran2016}. The optical gap in bulk hBN is on the order of 6~eV but the detailed nature of the VB to CB transition has been debated \cite{Cassabois:2016,Song2010,Majety2013,Ahmed2016}. 

The electronic dispersion and band gap of hBN depend on the detailed crystal structure and environmental conditions \cite{Topsakal:2009,Ribeiro2011,latini2015}. Indeed, $GW$ calculations of the quasiparticle bandstructure of AB stacked hBN predict an in-direct gap with values ranging from 5.4~eV \cite{Blase1995} to 5.95~eV in addition to a rich spectrum of tightly bound excitons \cite{Arnaud:2006}. However, the AB and AA$^{\prime}$ stacking orders where layers are staggered and eclipsed, respectively, placing B and N atoms on top of each other, are almost equally likely to occur \cite{Marom2010,Constantinescu2013}. This further complicates matters, as simple tight binding models have shown that these stacking orders are each characterized by a distinct electronic dispersion around the VB and CB extrema and therefore different band gap sizes \cite{Ribeiro2011}.

The electronic structure of hBN taking into account the crystal lattice potential, many-body effects and substrate interactions is best determined experimentally using angle-resolved photoemission spectroscopy (ARPES). However, photoemission measurements from insulating samples such as bulk hBN are typically excluded due to severe charging effects. A possible work-around to this problem involves growing single-layer hBN through borazine (B$_3$H$_6$N$_3$) decomposition on metal surfaces such as Ni(111) \cite{Nagashima:1996}, Ir(111) \cite{Usachov:2012,Orlando:2012} and Cu(111) \cite{Roth2013}. In these cases, hybridization between the hBN $\pi$-states and the electronic bands of the underlying metal substrate may significantly influence the bare hBN dispersion, requiring an intercalation step to decouple the hBN from the substrate \cite{Usachov2010,Verbitskiy2015,Fedorov:2015}. The metal substrates can be avoided altogether by directly growing the hBN on silicon carbide (SiC) followed by annealing above 1150~$^{\circ}$C, which produces a G/hBN heterostructure \cite{Shin2015}. 

The dominant role of the substrates in these studies makes a direct comparisson of electronic structures difficult between epitaxial hBN and the exfoliated thicker hBN flakes that are normally used for devices \cite{wang:2017}. In this work, we use the recently discovered trick that charging effects in photoemission from a bulk-like flake of an insulating material can be avoided by thinning the material in the exfoliation process and placing it on a conductive support \cite{Ulstrup:2016,Henck2017,Katoch:2018}. Indeed, a recent photoemission study showed that it was possible to resolve the $\pi$-band from hBN exfoliated on graphene \cite{Henck2017}. Here, we use micro-focused ARPES (microARPES) to completely determine the dispersion of the $\pi$- and $\sigma$-states over the full spectral energy range of these VBs. We use horizontal and vertical polarization of the photon beam to clearly separate the signals of the $\pi$- and $\sigma$-bands and explore their bulk $k_z$-dispersion via photon energy dependent measurements. Our approach reveals that an arbitrary conductive substrate such as degenerately doped TiO$_2$ can be used as a support for exfoliated flakes of hBN in order to collect photoemission spectra from flakes with lateral sizes reaching 200~$\mu$m. The measured hBN dispersion is compared with a tight binding model that accounts for the hBN stacking orders. We determine the VB bandstructure for both pristine and potassium-doped hBN. 

We investigate how the electronic properties of epitaxial hBN grown \textit{in situ} on graphene-covered silicon carbide (G/SiC) differ from the exfoliated bulk-like flakes. The band alignments and bandwidths of the different types of samples are measured using microARPES for the exfoliated flakes and $k$-resolved photoemission electron microscopy (PEEM) for the epitaxial samples. PEEM provides direct chemical and structural selectivity which permits us to study the hBN growth on graphene in real time and subsequently select a distinct micron-sized area for mapping the epitaxial hBN bandstructure. While the energy-resolution in the $k$-resolved PEEM mode is inferior to the microARPES, the simultaneous chemical, structural and electronic selectivity was critical for understanding the epitaxial hBN electronic structure. We find that the VB bandwidths and band separations of epitaxial and exfoliated hBN differ substantially, on the scale of electron volts.

\section{Experiments}
Flakes of hBN were exfoliated from a bulk crystal (hqgraphene) using residue-free scotch tape and placed onto a 0.5~wt~\% Nb-doped rutile TiO$_2$(100) wafer (Shinkosha Co., Ltd.). The sample was annealed for 1~hour at 625~K in ultrahigh vacuum (UHV) to remove tape residue. The transferred hBN flakes were optically inspected with a Nikon measuring microscope MM-40 before being inserted in the microARPES UHV end-station at the Microscopic and Electronic Structure Observatory (MAESTRO) at the Advanced Light Source (ALS). The sample was annealed at 625~K for 10~minutes in the end-station before measurements to remove adsorbates from the transfer of the sample through air. 

We investigated the electronic structure of flakes using a synchrotron beam with a lateral extent on the order of 10~$\mu$m, covering a photon energy range from 65 to 160~eV and using $p$- or $s$-polarized light. Photoemission spectra were collected from a single spot on a selected flake without moving the sample using a Scienta R4000 electron analyzer. Complete band mapping of the hBN Brillouin zone (BZ) from such a spot was carried out using custom-made deflectors in the analyzer, enabling the detection of photoelectrons from a cone of up to 30 degrees opening angle. Potassium was deposited \textit{in situ} on the hBN flakes from a SAES getter source mounted directly below the electron analyzer. The sample temperature was kept at 25~K during microARPES measurements using liquid helium cooling. The energy and momentum resolution were better than 20 meV and 0.01~\AA$^{-1}$, respectively.

Epitaxial growth of hBN was carried out \textit{in situ} on the graphene covered Si-terminated face of 6H-SiC(0001) in the SPECS FE-LEEM/PEEM P90 system installed at MAESTRO. Prior to growth, we used a Hg arc discharge lamp with an ultraviolet spectrum peaked at $\approx$4.5~eV to image the work function contrast on the SiC substrate to identify domains of the $6\sqrt{3}$-layer, monolayer graphene and bilayer graphene on the SiC surface. Borazine gas at a pressure of $1 \times 10^{-5}$~mbar was then dosed from a nozzle placed $\approx 1$~cm away from the sample surface. When the borazine pressure stabilized we ramped the temperature from 300~K to 1300~K, as measured with a pyrometer mounted in the line of sight of the sample surface in the PEEM measurement position. PEEM images were continuously collected from the sample during the entire process using the contrast from secondary electrons at the boron K-edge $\pi^{\ast}$-peak. This was achieved using synchrotron radiation with a photon energy of 192.2~eV and by filtering the secondary electrons with a contrast aperture in the diffraction plane of the microscope. The growth process was terminated once the signal of the $\pi^{\ast}$-peak saturated. 

$k$-resolved PEEM spectra were collected using a photon energy of 50~eV when the sample temperature had reached 300 K. The $k$-resolved measurements were performed using an aperture with a field of view (FOV) of 7.4~$\mu$m on the sample. The energy filtering was achieved using a magnetic prism with a motorized 1~$\mu$m entrance slit \cite{Fujikawa2009,Tromp2009}, which provided a total energy and momentum resolution on the order of 250~meV and 0.03~\AA$^{-1}$, respectively. All PEEM data were collected with an extraction voltage of -5~kV applied to the sample and keeping the objective lens 1.5~mm away from the sample and at ground potential.

\section{Results and Discussion}

\subsection{Electronic structure of exfoliated hBN determined by microARPES}

\begin{figure}
\begin{center}
\includegraphics[width=0.49\textwidth]{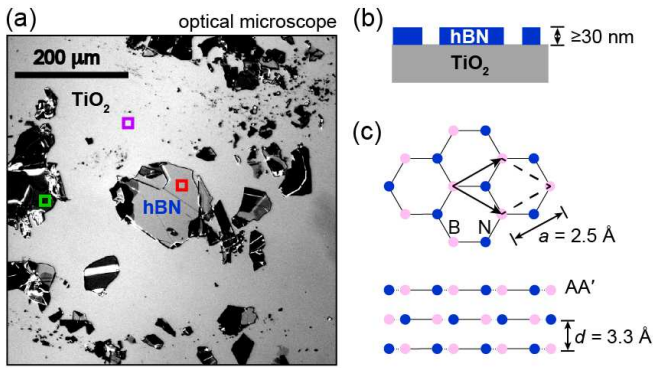}
\caption{(a) Optical microscope image of exfoliated hBN flakes on a TiO$_2$ substrate. The boxed regions display areas of the sample selected for the dispersion plots in Fig. \ref{fig:2}. (b) Sketch of the sample and indication of the order of magnitude of the hBN thickness. (c) Diagram of hBN crystal structure and the AA$^{\prime}$ stacking order identified from the photoemission data in Fig. \ref{fig:3}. The primitive unit cell and vectors are indicated by dashed lines and arrows, respectively. The lattice vector length $a$ and interlayer distance $d$ are provided.}
\label{fig:1}
\end{center}
\end{figure}

Our mechanically exfoliated hBN on TiO$_2$ yields flakes up to 200~$\mu$m wide as seen in the optical micrograph shown in Fig. \ref{fig:1}(a). The optical contrast levels are consistent with typical bulk-like flakes that have thicknesses larger than 30~nm as sketched in panel~(b) \cite{Lee:2010}. Some flakes with different thicknesses are overlapping, which leads to the varying optical contrasts between flakes \cite{Novoselov10451}. While these flakes can be regarded as bulk-like hBN the thinning of layers in the exfoliation process and placement on a conductive substrate is critical to avoid charging in photoemission measurements. Some of the thickest hBN flakes indeed show signatures of charging such as rigid energy shifts of the hBN bands and broadening of the spectral linewidths. Fig. \ref{fig:1}(c) presents an overview of the honeycomb B-N crystal structure, providing the experimentally determined lattice vector length $a = 2.5$~\AA~and interlayer distance $d=3.3$~\AA~according to Ref. \citenum{Han2008}. As discussed further down, we determine the relevant stacking of BN layers to follow the AA$^{\prime}$ sequence indicated here.   

\begin{figure}
\begin{center}
\includegraphics[width=0.49\textwidth]{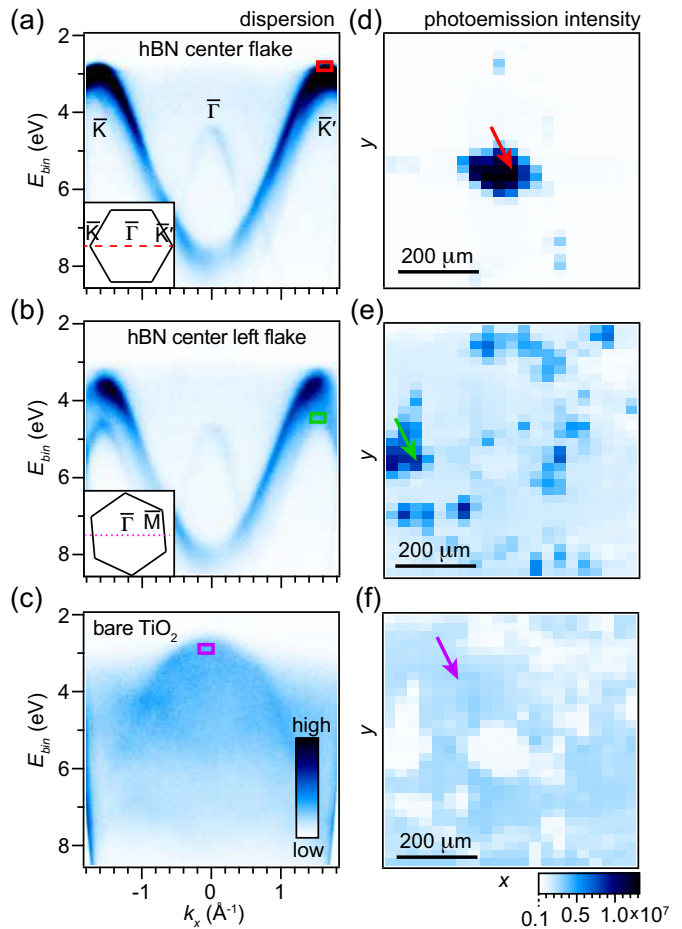}
\caption{(a)-(c) Area-selected microARPES measurements of the bandstructure of exfoliated hBN from the spots on the sample marked by arrows in the spatially distributed photoemission intensity plots in the corresponding row in (d)-(f). The dispersion is shown for (a) the center hBN flake where the \kgk~direction is along the cut (see BZ and dashed line in the inset), (b) the center left hBN flake which is oriented with the \mgm~direction close to the cut (see BZ and dotted line in the inset), and (c) a region of bare TiO$_2$ substrate. (d)-(f) Spatial intensity distributions obtained from the regions of $k$-space marked by rectangular boxes in (a)-(c).  The sample region measured here is the same as shown in Fig. \ref{fig:1}(a) where the areas indicated by arrows in (d)-(f) are marked by boxes.}
\label{fig:2}
\end{center}
\end{figure}

The microARPES spectra in Figs. \ref{fig:2}(a)-(c) provide the $E(k_x)$ dispersion relation averaged over a lateral extent of 10~$\mu$m from three distinct regions on the sample marked in the micrograph in Fig. \ref{fig:1}(a). These spectra were selected from a four-dimensional measurement of the $(E,k_x,x,y)$-dependent photoemission intensity as they provide representative snapshots of bulk hBN and surrounding TiO$_2$ bandstructures. We performed these measurements by scanning the sample position with a fixed photon beam over a 600$\times$600~$\mu$m$^2$ area in steps of 30~$\mu$m. We thus obtain a map of the photoemission intensity as a function of the $(x,y)$-coordinates where the intensity can be filtered within a specified region of $(E,k)$-space as shown in Figs. \ref{fig:2}(d)-(f). 

\begin{figure*} 
\begin{center}
\includegraphics[width=1\textwidth]{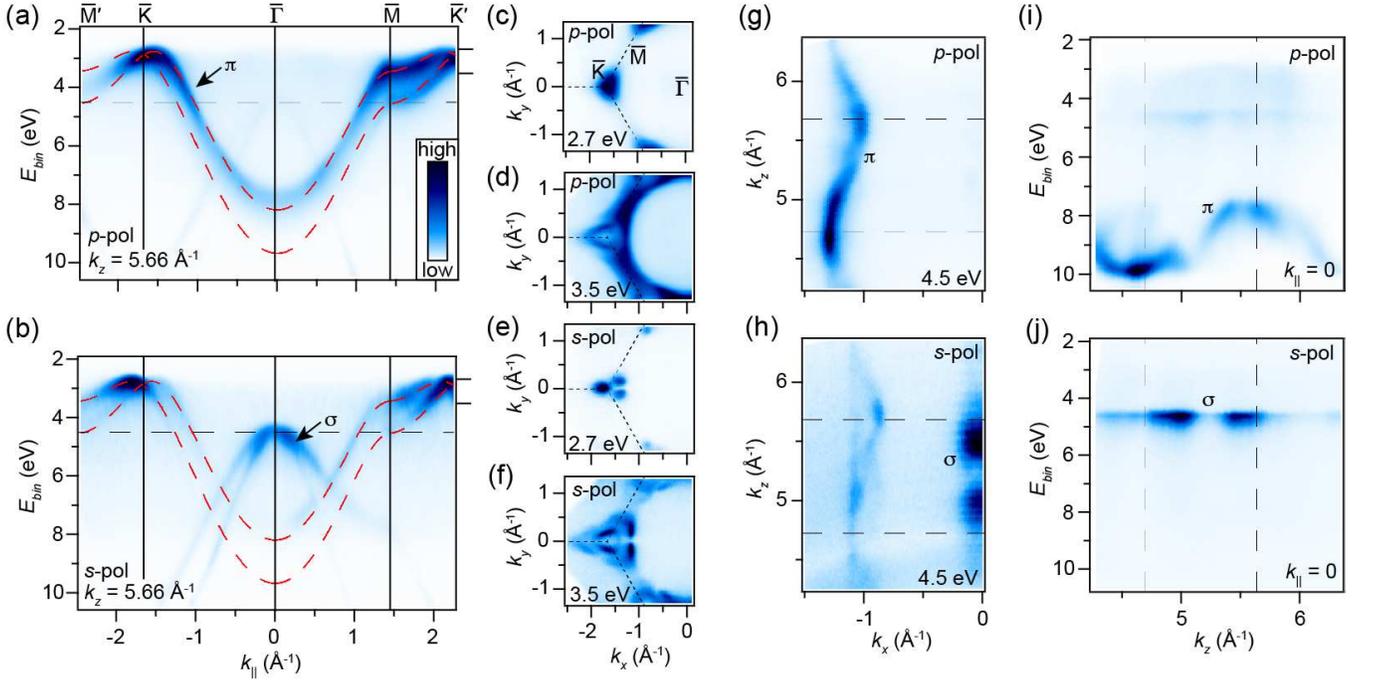}
\caption{(a)-(b) Measured $(E,k_{||})$-dispersion of hBN along the high symmetry directions of the BZ for a $k_z$ value of 5.66~\AA$^{-1}$. The $\pi$- and $\sigma$-bands of hBN are marked by arrows. The dashed dispersion curves are results of a tight binding model. (c)-(f) Constant ($k_x$,$k_y$)-energy surfaces at the given binding energies, which have been marked by ticks on the right side of panels (a)-(b). The dashed lines outline the BZ borders, and high symmetry points are marked in (c). (g)-(h) Constant ($k_x$,$k_z$)-energy surfaces at a binding energy of 4.5~eV exhibiting bulk dispersion of the $\pi$-band and intensity modulation at the top of the $\sigma$-band. (i)-(j) $(E,k_{z})$-dispersion of $\pi$- and $\sigma$-bands at $k_{||} = 0$. The dashed lines at fixed $k_{z}$ in (g)-(j) outline the BZ boundaries in the out-of-plane direction. The photoemission measurements in panels (a), (c)-(d), (g) and (i) were performed with $p$-polarized light probing mainly the $\pi$-bands, while the data in panels (b), (e)-(f), (h) and (j) were obtained with $s$-polarized light probing mainly the $\sigma$-bands. All datasets were collected from the center flake in Fig. \ref{fig:1}(a).}
\label{fig:3}
\end{center}
\end{figure*}

The center hBN flake in Fig. \ref{fig:1}(a) is oriented such that the $E(k_x)$-cut in Fig. \ref{fig:2}(a) provides the dispersion along the \kgk~direction of the hBN BZ (see inset in Fig. \ref{fig:2}(a)). The orientation of the BZ is determined through detailed $(E,k_x,k_y)$-dependent measurements of the intensity using the analyzer deflectors. The $(k_x,k_y)$-coordinates of the VB maxima at the \kbar~and \kbarp~points are determined using a constant energy contour corresponding to the maximum binding energy of the VB. Such contours are shown and discussed in further detail in connection with Figs. \ref{fig:3}(c)-(f) below. The hBN $\pi$-band is characterized by a strong photoemission intensity around the top of the band at \kbar. The intensity gradually diminishes and becomes diffuse towards \gbar, which signifies bulk-like dispersion of the $\pi$-band. The spatial distribution of the intensity measured within the red box (see panel~(a)) at the top of the $\pi$-band at \kbar~is shown in panel~(d). The very low intensity of the surrounding flakes is caused by the lack of states in this $E(k_x)$ region due to different crystal orientations. Plotting the intensity from the green box on the lower band at \mbar~in panel~(b) reveals the surrounding flakes in this region in the $(x,y)$-map in panel~(e). The two distinct bands around the top of \mbar~in panel~(b) derive from the two B-N layers in the unit cell of bulk hBN. The spatial intensity distribution in panel~(f) from the box placed on the characteristic TiO$_2$ feature in panel~(c) is characterized by an opposite contrast on the hBN flakes and surrounding TiO$_2$ compared to panels~(d)-(e), which were composed from hBN features. 

These microARPES data only represent a small subset of the full $(E,k_x,x,y)$-dependent photoemission intensity, but they demonstrate that we can distinguish the dispersion between hBN flakes with different orientations and between hBN and the TiO$_2$ substrate. In the following, we have collected all the photoemission data from the center hBN flake. 

The dispersion measurements along the high symmetry directions of the hBN BZ in Fig. \ref{fig:3}(a) reveal mainly the $\pi$-states that are probed in the $p$-polarized photoemission configuration here. Tuning the light to $s$-polarization leads to stronger sensitivity towards the in-plane $\sigma$-states, revealing the top of the $\sigma$-band located at a binding energy of 4.5~eV, as seen in Fig. \ref{fig:3}(b). A close inspection of the constant energy contours around the top of the $\pi$-band for the two polarizations reveal remarkably strong matrix element effects in the measured photoemission intensity \cite{Moser2017}. These are a consequence of interference effects between photoelectrons emitted from different sites within and between the BN layers. In Fig. \ref{fig:3}(c)-(d) the $\pi$-band evolves from a filled triangular contour to a triangular pocket connected by an arc that traverses the entire 1st BZ. The $s$-polarized measurement geometry instead reveals two sets of separated bands that cross at the \kbar-point (see panel (b)). The VB maximum is therefore located slightly away from \kbar, which leads to the three spots of intensity that surround the \kbar-point in Fig. \ref{fig:3}(e) and further produces a complex shape of the $\pi$-band contours at higher binding energies as seen in Fig. \ref{fig:3}(f).

The origin of the multiple $\pi$-band structure and degeneracy at \kbar~can be understood from the minimal tight binding models of bilayer hBN presented in Ref. \citenum{Ribeiro2011}. Following their calculations, the tight binding $\pi$-band structure of AA$^{\prime}$ stacked hBN is given by
\begin{eqnarray}
E = -\frac{1}{2}\sqrt{\Delta E^2 + 4\left({t^{\prime}}^2 + t^2|\phi |^2\right) \pm 8tt^{\prime}|\phi |},
\label{eq:1}
\end{eqnarray}
where 
\begin{eqnarray}
|\phi |^2 =  3 &+& 2\cos\left(\frac{- ak_x + \sqrt{3}ak_y}{2}\right)  + 2\cos\left(\frac{ak_x + \sqrt{3}ak_y}{2}\right) \nonumber \\
 &+& 2\cos\left(ak_x\right).
\label{eq:2}
\end{eqnarray}
Here, $\Delta E = E_B - E_N$ is the difference in the boron and nitrogen onsite energies, $t$ describes in-plane hopping, $t^{\prime}$ describes hopping between planes and $a = 2.5$~\AA~is the lattice vector length defined in Fig. \ref{fig:1}(c). Fitting this model to the measured dispersion yields $\Delta E = 5.53$~eV, $t = 2.83$~eV and $t^{\prime} = 0.78$~eV, leading to the dashed dispersion lines plotted on the photoemission intensity in Figs. \ref{fig:3}(a)-(b). the AB stacking model instead predicts that the top of the band is characterized by two bands that never cross but are separated in binding energy \cite{Ribeiro2011}. This clearly does not reproduce the degeneracy at \kbar~that is observed here and that the AA$^{\prime}$ model captures.

While the model reproduces the essential features around \kbar~we stress that our measurements are further complicated by the bulk $k_z$-dispersion, which leads to an intensity modulation of the $\pi$-band within the envelope of the tight binding dispersion lines, as seen in Fig. \ref{fig:3}(g)-(i). The $k_z$-values are obtained under the assumption of a free electron final state model where $k_z = 2m/\hbar^2\sqrt{E_k + V_0}$. Here, $E_k$ is the measured photoelectron kinetic energy and $V_0$ is the inner potential, which was set to 10~eV in order to obtain a meaningful description of the $k_z$-dispersion along the normal emission line. The $k_z$-dispersion of the minimum of the $\pi$-band in Fig. \ref{fig:3}(i) reveals photoemission intensity alternating between 9.7~and 7.9~eV. An apparent lattice doubling occurs in the $k_z$-direction as the periodicity of the $k_z$-dependent intensity seems to be closer to $2\pi/d$ rather than the unit cell length of $2\pi/2d$ where $d = 3.3$~\AA~is the interlayer distance defined in Fig. \ref{fig:1}(c). Similar behavior was observed in photoemission from multilayer stacks of graphene \cite{Ohta:2017}. The $k_z$-dispersion of the top of the $\sigma$-band in Fig. \ref{fig:3}(j) is completely flat, as expected for such states that derive from the sp$^2$ orbitals localized within the B-N planes.

\begin{figure}
\begin{center}
\includegraphics[width=0.49\textwidth]{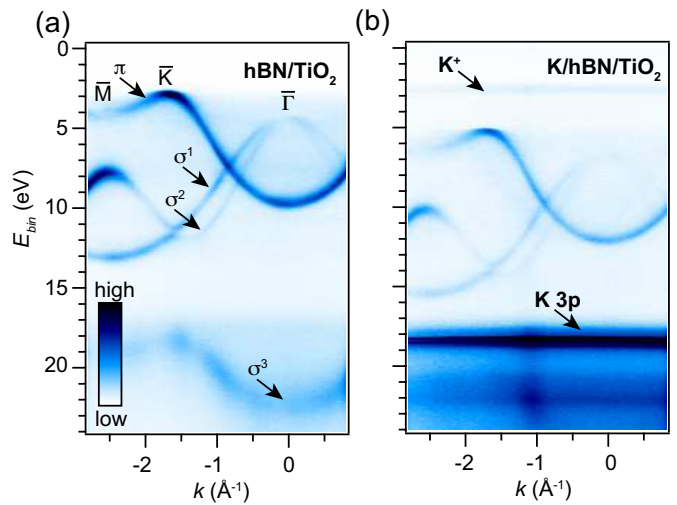}
\caption{(a) Wide energy range VB spectrum of hBN. The $\pi$- and $\sigma$-bands are marked by arrows. (b) Similar spectrum for potassium dosed hBN. The K 3p core level and the ionized K$^+$ line are marked by arrows. The data were obtained at a photon energy of 160~eV.}
\label{fig:4}
\end{center}
\end{figure}

The microARPES measurement along \gkm~in Fig. \ref{fig:4}(a) provides an overview of the dispersion covering the full VB energy range. The photon energy of 160~eV corresponds to a $k_z$-value where the $\pi$-band dispersion at \gbar~has a minimum (around $k_z = 6.5$~\AA$^{-1}$ in Fig. \ref{fig:3}(i)), which permits us to extract the full $\pi$-bandwidth of bulk hBN. We identify four subbands $\pi$, $\sigma^1$, $\sigma^2$ and $\sigma^3$ that are marked by arrows in panel~(a). The lifetimes of the $\sigma^3$ states are much lower than the states in the other bands because of the high binding energies, which causes the broad spectral linewidths.

Potassium (K) doping leads to a rigid shift of the $\pi$, $\sigma^1$ and $\sigma^2$ bands. The $\sigma^3$ band is obscured in our measurement because the strong K~3p state coincides with this binding energy region (see panel~(b)). We also observe a non-dispersive state around a binding energy of 2.5~eV which we attribute to the ionized potassium (K$^+$) that is adsorbed on the sample surface. The very high K doping (evidenced by the strong K~3p state) does unfortunately not provide access to the CB states of hBN.

\subsection{Epitaxial growth and bandstructure of hBN on G/SiC investigated by PEEM}

In order to establish the conditions for the hBN growth we start by determining the morphology of the G/SiC substrate. Monolayer graphene (MLG) is normally grown on SiC by thermal decomposition and removal of silicon \cite{Forbeaux:1998}. This process does typically not lead to a homogeneous MLG coverage but a mix of MLG, bilayer graphene (BLG) and the $6\sqrt{3}$-layer (or buffer layer) that passivates the dangling bonds in the SiC substrate \cite{Emtsev:2009,Ostler:2010}.  

\begin{figure} 
\begin{center}
\includegraphics[width=0.4\textwidth]{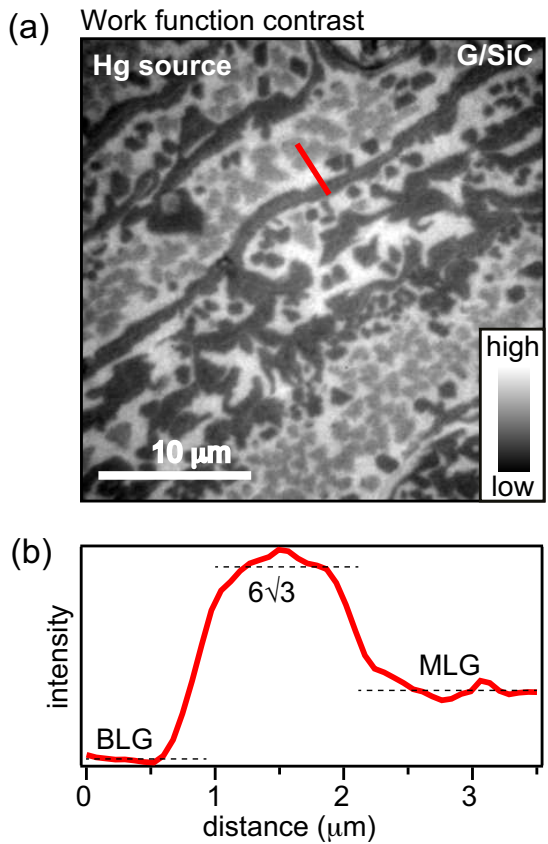}
\caption{(a) PEEM image of as-grown epitaxial graphene on silicon carbide obtained with a Hg lamp probing the work function contrast. (b) Intensity contrast along the line in (a), showing three contrast levels relating to the $6\sqrt{3}$-layer, monolayer graphene (MLG) and bilayer graphene (BLG).}
\label{fig:5}
\end{center}
\end{figure}

In Fig. \ref{fig:5}(a) we show a PEEM image of a representative region of our G/SiC substrate, which was acquired using 4.5~eV photons from a Hg source. The measurement is therefore highly sensitive to work function differences across the sample surface. The intensity profile in Fig. \ref{fig:5}(b) was taken across the red line in panel~(a) and shows that three distinct contrast levels are present in the image. Since the work function increases by $\approx 300$~meV from the $6\sqrt{3}$-layer to graphite we can attribute the decreasing contrast level between patches to an increasing number of graphene layers \cite{Hibino2009,Hibino2016}, as noted in panel~(b).  

\begin{figure}
\begin{center}
\includegraphics[width=0.49\textwidth]{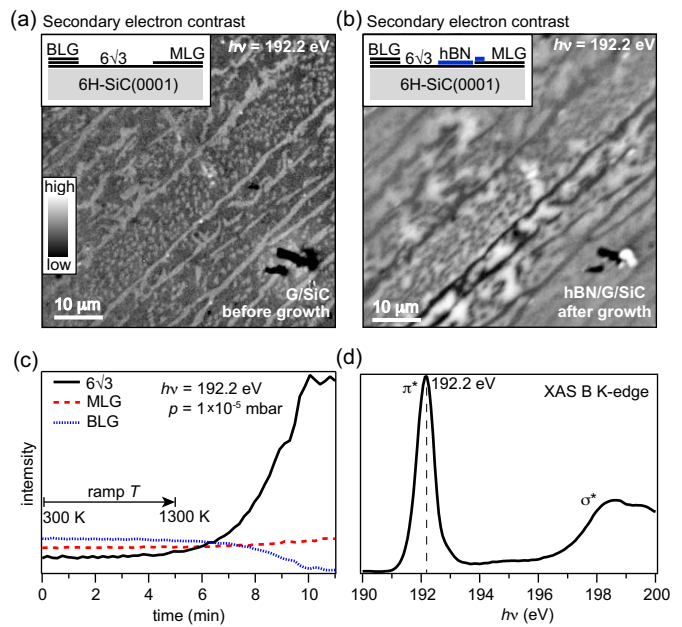}
\caption{(a)-(b) PEEM images obtained with secondary electron contrast using synchrotron radiation at a photon energy of 192.2~eV (a) before and (b) after hBN growth (see insets for sketches of the sample under the given conditions). (c) Photoemission intensity from panels (a)-(b) summed over the three distinct regions of the substrate as a function of time during the hBN growth process. The substrate temperature is ramped from 300~K to 1300~K during the first 5~min and the borazine pressure is kept at $1\times 10^{-5}$~mbar in the entire process. (d) XAS spectrum of the B K-edge obtained by scanning the PEEM secondary electron contrast of the grown hBN/G/SiC film over the given photon energy range, revealing the characteristic $\pi^{\ast}$- and $\sigma^{\ast}$-peaks. The dashed line marks the feature used for obtaining the data in panels (a)-(b).}
\label{fig:6}
\end{center}
\end{figure}

The growing hBN layer is most efficiently monitored in PEEM using the boron K-absorption edge which has a distinct peak at a photon energy of 192.2~eV that corresponds to the $\pi^{\ast}$-resonance. We therefore proceed by recording PEEM images of the hBN growth while irradiating photons at this energy from the synchrotron. Fig. \ref{fig:6}(a) presents a PEEM image of the pristine G/SiC surface using secondary electron contrast at 192.2~eV. The sample region is the same as in Fig. \ref{fig:5}(a) but a wider FOV is shown, and the three contrast levels have reversed compared to the work function contrast. The situation after growth, in Fig. \ref{fig:6}(b), is characterized by high contrast levels on the $6\sqrt{3}$ plateaus of the substrate. 

The full growth process on the different graphene regions is tracked by integrating the $\pi^{\ast}$-peak intensity separately on the $6\sqrt{3}$, MLG and BLG regions for each image acquired during the growth \cite{suppl}, see Fig. \ref{fig:6}(c). The borazine pressure is always kept at $1 \times 10^{-5}$~mbar, but it does not lead to BN deposition on the G/SiC before the temperature reaches 1300~K. Once at the maximum temperature the hBN nucleates and grows on the $6\sqrt{3}$-layer. A slight $\pi^{\ast}$-peak intensity increase is also observed on MLG parts, which we interpret as hBN layers expanding across these areas from the hBN nucleation centers on the $6\sqrt{3}$ parts (see insert in Fig. \ref{fig:6}(b)). On the BLG areas the interaction with the growing hBN is too weak to establish a continuous film. We can not exclude that these areas are becoming more disordered during growth, which would explain the surprising intensity decrease on the BLG parts. 

Fig. \ref{fig:6}(d) presents an x-ray absorption spectrum (XAS), which was determined by measuring the secondary electron contrast on the sample area seen in panels~(a)-(b) while changing the photon energy. The characteristic $\pi^{\ast}$- and $\sigma^{\ast}$-resonances ascertain that the grown film has the electron configuration expected for hBN \cite{Usachov2010}.

The process observed here shares many similarities with the van der Waals epitaxy of other 2D materials such as MoS$_2$ on G/SiC \cite{Miwa:2015}, where nucleation also starts on the $6\sqrt{3}$ parts and growth extends onto MLG areas. 

\begin{figure}
\begin{center}
\includegraphics[width=0.49\textwidth]{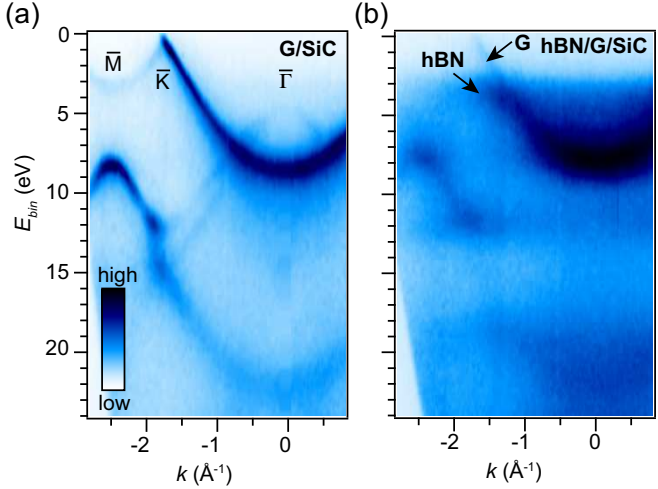}
\caption{(a)-(b) $k$-resolved PEEM measurements of the full VB bandwidth obtained at a photon energy of 50~eV (a) before and (b) after hBN growth on graphene on silicon carbide. Features relating to the hBN and graphene (G) $\pi$-bands are marked by arrows in (b).}
\label{fig:7}
\end{center}
\end{figure}

The $k$-resolved PEEM measurements of the bandstructure from a 7.4~$\mu$m area on the sample before and after growth are shown in Figs. \ref{fig:7}(a)-(b). We here focus on a sample area that has $6\sqrt{3}$ and MLG graphene, which is fully covered by hBN after growth. The situation in panel~(a) before growth displays the typical dispersion of the $\pi$- and $\sigma$-bands of MLG/SiC with a very slight $n$-doping that is difficult to resolve in these measurements \cite{Emtsev:2009}. After growth the dispersion is instead dominated by hBN features, which are characterized by a VB maximum towards \kbar~at higher binding energies (see hBN arrow in panel~(b)), a smaller bandwidth of the $\pi$-band compared to graphene and a large gap between the high binding energy $\sigma$-states (between 13 and 17~eV). These features resemble those of the hBN/TiO$_2$ in Fig. \ref{fig:4}(a).

We observe the top of the MLG $\pi$-states under the hBN (see G arrow in Fig. \ref{fig:7}(b)). In these PEEM measurements we are only sensitive to the top-most layers of the sample since we are probing the photoemitted electrons using 50~eV photons. The intensity from the MLG bandstructure suggests that the hBN has monolayer thickness, as also found in Ref. \citenum{Shin2015} using a similar approach for the growth. Further measurements of the full hBN bandstructure (not presented) within the 7.4~$\mu$m FOV show that hBN lacks azimuthal order. The very weak interaction with the substrate leads to the growth of nano-sized hBN islands with arbitrary orientation, similarly as seen for other van der Waals materials on G/SiC \cite{Miwa:2015}. 

\subsection{Analysis of hBN band alignments}

The hBN/G/SiC and hBN/TiO$_2$ systems both display the main $\pi$-, $\sigma^1$-, $\sigma^2$- and $\sigma^3$-bands identified in Fig. \ref{fig:4}(a). For hBN/G/SiC we do not observe the split $\pi$-band structure as seen for the thick hBN flakes in Figs. \ref{fig:2} and \ref{fig:3}, presumably due to the monolayer thickness. The most dramatic differences between the samples emerge from the binding energies and bandwidths of the $\pi$- and  $\sigma^3$-bands, which are compared in detail in the following.

\begin{figure}
\begin{center}
\includegraphics[width=0.49\textwidth]{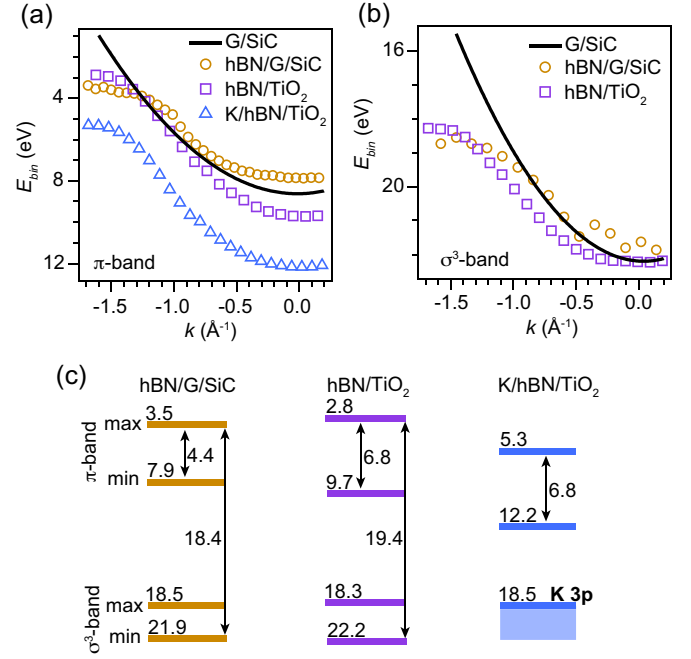}
\caption{(a)-(b) Peak positions obtained from EDC analysis of the (a) $\pi$-band and (b) $\sigma^3$-band in the spectra in Figs. \ref{fig:4} and \ref{fig:7}. (c) Band separations and total VB bandwidth for the different hBN samples. Energies are stated in units of eV.}
\label{fig:8}
\end{center}
\end{figure}

We can quantitatively compare the peak positions of energy distribution curves (EDCs) of the $\pi$- and $\sigma^3$-bands of exfoliated and epitaxial hBN obtained with microARPES and PEEM, respectively. This approach is advantageous over spectral linewidth analysis as it is not significantly affected by the difference in energy resolution and overall background signal. In order to directly compare bandwidths and band separations we therefore extract the dispersion of the  $\pi$- and $\sigma^3$-bands using EDC peak position analysis as shown in Figs. \ref{fig:8}(a)-(b). The corresponding graphene dispersion is shown as a reference in each case. Bandwidths and binding energy positions are summarized in the diagrams for the three different hBN samples (hBN/G/SiC, hBN/TiO$_2$ and K/hBN/TiO$_2$) in panel~(c). 

The different VB binding energy positions between hBN/G/SiC and hBN/TiO$_2$  can be explained by the different Fermi level pinning in these systems. In the hBN flakes the pinning will be defined by the bulk defects while on G/SiC it would depend on the band alignments between the different materials. The apparent 2.4~eV reduction in the bandwidth between hBN/G/SiC and hBN/TiO$_2$ suggests that the bandstructure of epitaxial hBN is modified by the substrate, possibly due to interaction with the $6\sqrt{3}$ or epitaxial strain as discussed in Ref. \citenum{Shin2015}. The bandwidth may also be influenced by the islanded morphology of the epitaxial hBN, which may cause scattering and localization effects. Quasi free-standing monolayer hBN achieved by intercalation in Refs. \citenum{Usachov2010,Verbitskiy2015}~ appears to have a $\pi$-bandwidth that is closer to the value of 6.8~eV we obtain for the bulk hBN. Note that the VB bandwidths reported for hBN/G/SiC in panel~(c) may also be underestimated due to the lack of azimuthal order. This leads to a \kbar~point that has broad energy- and momentum-distributions. The uncertainty in the determination of the binding energy position is therefore larger than for the other samples. We do not observe any signs of hybridization effects between the hBN and graphene bands in these data.

The full VB bandwidth of 19.4~eV for bulk hBN is in line with the $GW$ calculation for AB stacked hBN that provided a value of 19.87~eV \cite{Blase1995}. We can not determine the quasiparticle band gap since the K-dosing does not reveal the CB states, but we can determine that it is larger than the VB maximum of 5.3~eV in the fully doped case. It may be possible to dope the CB states of a single-layer of hBN by a combination of intercalation and adsorption of two strongly $n$-doping atoms such as potassium and calcium, as previously demonstrated for graphene \cite{McChesney:2010}. Such strong $n$-doping may provide new insights in the CB dispersion, but it may also significantly change the quasiparticle excitations around the VB and CB extrema due to doping induced many-body effects \cite{Katoch:2018}. 

\section{Conclusion}

We have measured the $(k_x,k_y,k_z)$-dependence of the electronic dispersion in exfoliated flakes of hBN using microARPES. The dispersion could be fitted with a minimal tight binding model for AA$^{\prime}$ stacked hBN where the VB maximum is located slightly away from the \kbar~point. Monitoring of epitaxial growth of hBN on the $6\sqrt{3}$ and MLG parts of G/SiC was achieved \textit{in situ} by tracking the secondary electron contrast of the $\pi^{\ast}$-resonance using PEEM during thermal decomposition of borazine. The epitaxial hBN had a significantly reduced bandwidth compared to the exfoliated thicker hBN flakes where the full VB electronic structure more closely resembled the theoretical prediction from \textit{GW} calculations. Potassium doping of exfoliated hBN lead to a rigid shift of the VB with the band edge moving from a binding energy of 2.8~ to 5.3~eV while the CB remained unoccupied.

\section{acknowledgement}
R. J. K. is supported by a fellowship within the Postdoc-Program of the German Academic Exchange Service (DAAD). S. U. acknowledges financial support from the Danish Council for Independent Research, Natural Sciences under the Sapere Aude program (Grant No. DFF-4090-00125) and from VILLUM FONDEN (Grant. No. 15375). S. M. acknowledges support by the Swiss National Science Foundation (Grant No. P300P2-171221). D. S. acknowledges financial support from the Netherlands Organisation for Scientific Research under the Rubicon Program (Grant 680-50-1305). The Advanced Light Source is supported by the Director, Office of Science, Office of Basic Energy Sciences, of the U.S. Department of Energy under Contract No. DE-AC02-05CH11231. This work was supported by IBS-R009-D1. The work at Ohio State was primarily supported by NSF-MRSEC (Grant DMR-1420451).

\end{document}